\documentclass[twocolumn]{autart} 

\usepackage{amsmath,amssymb}
\usepackage{color}

\newcommand{\R}{\mathbb R}
\newcommand{\Rp}{\mathbb R_{\geq 0}}
\newcommand{\cP}{\mathcal P}
\newcommand{\bP}{\mathbb P}
\newcommand{\cD}{\mathcal D}

\newcommand{\N}{\mathbb N}

\DeclareMathOperator{\re}{\textnormal{Re}}

\DeclareMathOperator*{\ramp}{\textnormal{ramp}}

\allowdisplaybreaks

\newcommand{\revise}[1]{{\color{black}#1}}

\begin{document}

\begin{frontmatter}

\title{Unified stability criteria for perturbed {LTV} systems with unstable instantaneous dynamics}


\author[a]{Shenyu Liu}\ead{shl055@ucsd.edu},    
\address[a]{Mechanical and Aerospace Engineering, University of California, San Diego, USA}  


\begin{abstract}                          
In this work the stability of perturbed linear time-varying systems is studied. The main features of the problem are threefold. Firstly, the time-varying dynamics is not required to be continuous but allowed to have jumps. Also the system matrix is not assumed to be always Hurwitz. In addition, there is nonlinear time-varying perturbation which may be persistent. We first propose several mild regularity assumptions, under which the total variations of the system matrix and its abscissa are well-defined over arbitrary time interval. We then state our main result of the work, which requires the combined assessment of the total variation of the system matrix, the measure when the system is not sufficiently ``stable" and the estimate of the perturbation to be upper bounded by a function affine in time. When this condition is met, we prove that the neighborhood of the origin, whose size depends on the magnitude of the perturbation, is uniformly globally exponentially stable for the system.
We make several remarks, connecting our results with the known stability theory from continuous linear time-varying systems and switched systems. Finally, a numerical example is included to further illustrate the application of the main result.
\end{abstract}

\end{frontmatter}

\section{Introduction}
Due to the long-lasting importance of the design and analysis of adaptive controllers, 
the stability analysis for linear time-varying (LTV) systems has played an important role in control theory for decades \cite{Ioannou2006}. The early study of stability of LTV systems can date back to the work \cite{Desoer1969}. Since then it is well-known that even if the instantaneous dynamics of the system is stable and the abscissa of the system matrix is uniformly upper-bounded by some negative number, an LTV system may still be unstable. In order to ensure global asymptotic stability, one needs the system to vary ``slowly", in the sense that either the time derivative of the system matrix has sufficiently small magnitude \cite{Coppel1978,Amato1993}, or the variation of the system matrix is upper-bounded on average \cite{ILCHMANN1987157}. When the instantaneous dynamics is not necessarily always stable, the work \cite{Solo1994,Jetto09} propose different sets of conditions under which the LTV systems are exponentially stable. However, these results have fairly complex assumptions and hence can not be easily applied to real problems. 
Meanwhile, the works \cite{ZHOU2016266,CHEN2016342} based on indefinite Laypunov function have nicer results which may be useful for concluding stability of LTV systems with possible unstable instantaneous dynamics.

Since the end of the 20th century, the study of switched systems has gradually gain its importance in control theory because of its wide application in modern engineering problems \cite{Liberzon2003b}. Switched systems are essentially a special class of time-varying systems, whose dynamics varies in a piece-wise continuous manner. Similar to the feature of time-varying systems, stability of a switched system is not guaranteed either even if all its modes are stable. Researchers hence developed different criteria such as dwell-time condition, average dwell-time condition \cite{Morse1993,Hespanha1999} which bound the number of switches over an arbitrary time interval. Stability of a switched system can then be shown when its switching signal satisfies these criteria. When some modes of a switched system is unstable, criteria on the switching signal which bound the average activation time of the unstable modes can be used for proving its stability in \cite{Zhai2001,Mueller2012}. A similar approach of using indefinite Lyapunov function is also used in \cite{Long2019} to study the stability of switched time-varying systems.

On the other hand, because of uncertainties, linearization, modeling error, external disturbance or other perturbation factors, no dynamical system is truly linear with the exact system matrix in the real world \cite{Verhulst2009}. In this case, the dynamics of the true system can be modeled as the sum of a nominal LTV system and a perturbation term. While the nominal system can be shown stable via various stability analysis approaches,  different hypothesis are then imposed on the perturbation term in order to guarantee stability of the true system \cite{Vidyasagar2002,Khalil2002}. When the additive disturbance is treated as an external input, the equivalent characterizations of input-to-state stability for switched time-varying systems are studied in \cite{Haimovich2018,Haimovich2019}. However, in those works, no sufficient conditions on the switching signal as well as the time-varying nature are given for input-to-state stability. Recently, exponential stability of switched LTV systems with perturbations in the form of delays is studied in \cite{LI2017284}. In that work since the aimed stability property is uniform with respect to arbitrary switching, the concluded conditions for exponentially stable are conservative.

In this work we aim to study exponential stability of perturbed LTV systems. Compared with the aforementioned literature, the main features of this work are threefold. Firstly, the time-varying dynamics is not required to be continuous but allowed to have jumps. To the author's knowledge, such combination of continuous time-variation and switches are not studied together until the recent work \cite{Gao2018}, where a unified stability criteria based on total variation is proposed for such systems. \revise{Nevertheless, the system matrix is assumed to be piece-wise continuously differentiable in that work, whereas in this work we have a strictly weaker assumption that the system matrix is only assumed to be piece-wise absolutely continuous.} Moreover, the work \cite{Gao2018} does not allow the instantaneous dynamics to be unstable. This brings the second feature of our work that the system matrix is not assumed to be always Hurwitz. In addition, we assume the presence of nonlinear time-varying perturbation which may be persistent. Similar problems about stability of perturbed switched time-varying systems is studied in the works \cite{LIU2017114,mancillaaguilar2021integraliss}, where in the first work the perturbations are additive disturbance and delays, while in the second work the perturbations are impulses and errors due to linearization. In these works, the nominal systems need to be assumed uniformly exponentially stable, whereas in our work there is no such an assumption. Instead, in this work we propose a unified criteria based on the combined assessment of the total variation of the system matrix, the measure of the instantaneous dynamics when it is not sufficiently ``stable" and the estimate of the perturbation. The main contribution of this work is the conclusion that when this combined assessment is upper bounded by a function affine in time, then the neighborhood of the origin, whose size depends on the magnitude of the persistent perturbation, is uniformly globally exponentially stable. \revise{In terms of methodology, a Lyapunov-based approach is used to conclude the stability property. Similar Lyapunov-based approach also appears in \cite{Yang2015} for the stability analysis of interconnected switched systems. In that work, a Lyapunov function consists of an auxiliary timer is constructed. While we also use a timer in the construction of the Lyapunov function in order to make it monotonic along perturbation-free solution trajectories, because the system is time-varying and the system matrix is only assumed to be piece-wise absolutely continuous with respect to time, the regularity of the timer needs to be carefully discussed.}

The rest of the paper is organized as follows. Section~\ref{sec:prelim} gives the necessary notions and backgrounds for this work. Section~\ref{sec:A} discusses the assumptions on the LTV systems and some technical results we need for proving our main theorem. Section~\ref{sec:thm} then states the main theorem, followed with its proof. In Section~\ref{sec:example} we illustrate one numerical example on which our theorem can be applied to conclude its uniform global exponential stability. Finally Section~\ref{sec:conclusion} concludes the paper.

\section{Preliminaries}\label{sec:prelim}
Let $\R$ be the space of real numbers and $\mathbb C$ be the space of complex numbers. Let $\Rp:=[0,\infty)$ be the non-negative real line, $\mathbb Z$ be the space of integers and $\N:=\{1,2,\cdots\}$ be the set of natural numbers. For $A\in\R^{n\times n}$, let $\alpha(A)$ denote its spectral abscissa; that is,
\[
\alpha(A)=\max\{\re(\lambda):\lambda\in\mathbb C, \det(\lambda I-A)=0\}
\]
where $\re(\cdot)$ denotes the real part. The matrix $A$ is \emph{Hurwitz} if and only if $\alpha(A)<0$.

\revise{The \emph{weak derivative} of a real-valued function $f(t):[a,b]\mapsto\R$, denoted by $\dot f(t)$, is defined to be a function $g(t):[a,b]\mapsto\R$ such that
\[
\int_a^bf(t)\dot\varphi(t)dt=-\int_a^bg(t)\varphi(t)dt
\]
for all differentiable functions $\varphi(t):[a,b]\mapsto\R$ with $\varphi(a)=\varphi(b)=0$. When $f(t)$ is absolutely continuous over $[a,b]$, it follows from fundamental theorem of Lebesgue integral calculus that $f(b)-f(a)=\int_a^b\dot f(t)dt$. The weak derivative of a matrix-valued function $A(t):[a,b]\mapsto\R^{n\times n}$, denoted by $\dot A(t)$, is the matrix of the element-wise weak derivatives.}
For a vector $x\in\R^n$, we use $|x|$ to denote its 2-norm and for a matrix $A\in\R^{n\times n}$, we use $\Vert A\Vert$ to denote 2-norm induced norm. 
Given a matrix trajectory $A(t):\Rp\mapsto\R^{n\times n}$, its \emph{total variation} over an interval $[a,b]$ is denoted by $\int_{a}^{b}\Vert dA\Vert$, and is defined by
\begin{equation}\label{def:total_variation}
    \int_{a}^{b}\Vert dA\Vert:=\sup_{P\in\bP}\sum_{i=1}^{k}\Vert A(t_{i})-A(t_{i-1})\Vert
\end{equation}
where $P=\{t_0,t_1,\cdots,t_k\}$ with $t_0=a,t_k=b$ is a partition of $[a,b]$ and $\bP$ is the collection of all partitions of $[a,b]$. Notice that when the dimension of matrix $n=1$, the definition of total variation of a matrix trajectory coincides with the definition of total variation of a real-valued function.  
For any real-valued or matrix-valued function $f(t)$ whose left limit exists everywhere, denote $f(t^-):=\lim_{s\to t^-}f(s)$.

Consider a \emph{linear time-varying} (LTV) system with nonlinear state-dependent, time-varying perturbation
\begin{equation}\label{def:ltv_sys}
    \dot x(t)=A(t)x(t)+g(t,x)
\end{equation}
where $x(t)\in\R^n$ is the state, $g(t,x)\in\R^n$ is the perturbation and $A(t)\in\R^{n\times n}$. The regularity assumptions of the matrix trajectory $A(t)$ and perturbation $g(t,x)$ will be discussed in the next section. For a given initial state $x_0$ at time $t_0$, denote the solution of \eqref{def:ltv_sys} at time $t$ by $x(t;t_0,x_0)$ and when the initial pair $t_0,x_0$ is clear from the context, we use the abbreviation $x(t)$ instead. We say that the system \eqref{def:ltv_sys} has \emph{unstable instantaneous dynamics} at time $s$ if $A(s)$ is non-Hurwitz; i.e., the time-invariant nominal system $\dot x(t)=A(s)x(t)$ is unstable. 

\section{On the matrix trajectory $A(t)$ and perturbation $g(t,x)$}\label{sec:A}
In this work we would like to study the stability of the aforementioned LTV system with perturbation \eqref{def:ltv_sys} when the matrix trajectory $A(t):\Rp\mapsto\R^{n\times n}$ ``varies slowly" and the perturbation $g(t,x):\Rp\times\R^n\mapsto\R^n$ is ``small".
\subsection{Regularity assumptions}
We start with introducing two sets of regularity assumptions on $A(t)$. The first assumption ensures that $A(t)$ is in a compact set for all $t\in\Rp$:
\begin{assum}\label{ass:1}
    There exists $L>0$ such that $\Vert A(t)\Vert\leq L$ for all $t\in\R$.
\end{assum}
\revise{Notice that Assumption~\ref{ass:1} also implies the existence of $\alpha_{\max}\leq L$ such that $\alpha(A(t))\leq \alpha_{\max}$ for all $t\in\R$. However, we do not assume $\alpha_{\max}<0$;}  in other words, the system \eqref{def:ltv_sys} is allowed to have unstable instantaneous dynamics.  

The second set of assumptions describes how $A(t)$ and $\alpha(A(t))$ vary with respect to $t$:
\begin{assum}\label{ass:2}
Given any $b> a\geq 0$, the matrix trajectory $A(t):\Rp\to \R^{n\times n}$ satisfies
\begin{enumerate}\renewcommand{\labelenumi}{\arabic{enumi}.}
    \item $A(t)$ is a C\`adl\`ag function on $[a,b]$; i.e., it is right continuous and has left limit everywhere on $(a,b]$.
    \item $A(t)$ has finitely many discontinuities on $(a,b)$; i.e., denote
    \begin{equation}\label{def:D}
    \cD:=\{t\in\Rp:A(t)\neq A(t^-)\},
\end{equation}
then the set $\cD\cap(a,b)$
    has finite cardinality.
    \item Let $t_1,t_2,\cdots t_{p-1}$ be the elements of $\cD\cap(a,b)$ with the ordering that $a=:t_0<t_1<\cdots<t_{p-1}<t_p:=b$. $A(t)$ is absolutely continuous on $[t_i,t_{i+1})$ for all $i=0,1,\cdots,p-1$.
    \item $\alpha(A(t))$ is absolutely continuous on $[t_i,t_{i+1})$ for all $i=0,1,\cdots,p-1$, where $t_i$'s are the same as defined earlier.
\end{enumerate}
\end{assum}
Note that Assumption~\ref{ass:2}.1 and Assumption~\ref{ass:2}.2 allow $A(t)$ to jump occasionally. 
In addition, the piece-wise absolute continuity properties of $A(t)$ and $\alpha(A(t))$ in Assumption~\ref{ass:2}.3 and assumption~\ref{ass:2}.4 allow us to quantitatively characterize the ``slow variation" nature of $A(t)$, which will be discussed later in Section~\ref{subsec:slow_variation}.

\begin{rem}
Assumption~\ref{ass:2}.3 does not guarantee Assumption~\ref{ass:2}.4 in general. To see this, consider the example where
\begin{equation}
    A(t)=\begin{pmatrix}
    0& 1\\
    \mu(t)&0
    \end{pmatrix},
\end{equation}
where
 \[
\mu(t)=\left\{\begin{array}{cc}
    t^2\sin^2(\frac{1}{t}) &  \mbox{ if }t\neq 0,\\
    0 & \mbox{ if } t=0.
\end{array}\right.
\]
We first observe that $\Vert\dot A(t)\Vert=|\dot\mu(t)|$, and
\[
\dot\mu(t)=\left\{\begin{array}{cc}
    2s\sin^2(\frac{1}{t})-\sin(\frac{1}{t})\cos(\frac{1}{t}) & \mbox{ if } t\neq 0, \\
    0 & \mbox{ if }t=0.
\end{array}\right.
\]
Because the derivative exists everywhere on $[0,1]$ and it is bounded, $A(t)$ is absolutely continuous on $[0,1)$. However, we also observe that the eigenvalues of $A(t)$ are $\pm\sqrt{\mu(t)}$, so
\[
\alpha(A(t))=\begin{cases}
	|t\sin(\frac{1}{t})|&\mbox{ if }t>0,\\
	0&\mbox{ if }t=0
\end{cases}
\]
which is not absolutely continuous on $[0,1)$. \\
\revise{From the perspective of perturbation theory, This problem is caused by the arbitrarily huge sensitivity of eigenvalues when $A(t)$ is ``ill-posed" (Note in this example, $A(0)=\begin{pmatrix}
	0&1\\0&0
\end{pmatrix}$ is not diagonalizable). In other words, when $A(t)$ is absolutely continuous so that its variation is bounded, its abscissa can vary drastically and hence not absolutely continuous. Nevertheless, such problem can be avoided if the eigenvector matrix of $A(t)$ has uniformly bounded condition number. This result is stated by the next Lemma:}
\end{rem}
\revise{\begin{lem}
	Consider a matrix trajectory $A(t):\Rp\to \R^{n\times n}$ satisfying Assumption~\ref{ass:2}.3. If $A(t)$ is diagonalizable and there exists $k>0$ such that $\frac{\Vert V(t)\Vert}{\Vert V(t)^{-1}\Vert}\leq k$ for all $t\in[a,b]$, where $A(t)=V(t)\Lambda(t)V(t)^{-1}$ is the matrix diagonalization. Then Assumption~\ref{ass:2}.4 also holds.
\end{lem}
\begin{pf}
	Let $s,t\in[a,b]$ be arbitrary. Without loss of generality assume $\alpha(A(s))\geq\alpha(A(t))$ and denote a leading eigenvalue of $A(s)$ to be $\lambda_s$; that is, $\re(\lambda_s)=\alpha(A(s))$. We have $\alpha(A(s))-\alpha(A(t))\leq \alpha(A(s))-\re(\lambda)$, where $\lambda$ can be any eigenvalue of $A(t)$, including the one coming from Bauer–Fike theorem \cite{BF:60} satisfying
	\[
	|\lambda_s-\lambda|\leq \frac{\Vert V(t)\Vert}{\Vert V(t)^{-1}\Vert}\Vert A(s)-A(t) \Vert.
	\]
	Therefore
	\begin{multline*}
	|\alpha(A(s)-\alpha(A(t))|\leq |\alpha(A(s))-\re(\lambda)|\\
	\leq |\lambda_s-\lambda|\leq k\Vert A(s)-A(t) \Vert.
	\end{multline*}
	This lemma can then be proven by appealing to Lemma~\ref{lem:Lipschitz} in Appendix~\ref{sec:lem:ac}.
\end{pf}}

Finally, we have one assumption with respect to the perturbation:
\begin{assum}\label{ass:3}
The perturbation $g(t,x):\Rp\mapsto\R^n$ is Lebesgue integrable in $t$ for each fixed $x$, and locally Lipschitz in $x$ for each fixed $t$. Moreover, there are non-negative continuous functions $\gamma,\delta:\Rp\mapsto\Rp$ such that
\begin{equation}\label{bound_on_g}
|g(t,x)|\leq \gamma(t)|x|+\delta(t)\quad\forall (t,x)\in\Rp\times\R^n.
\end{equation}
\end{assum}

The inequality \eqref{bound_on_g} is a standard assumption on the perturbation (cf., \cite[Equation~(9.15)]{Khalil2002}). When $\delta(t)\equiv 0$, the perturbation is \emph{vanishing} since the magnitude of the perturbation decreases to $0$ when $x$ approaches to the origin. Stable unmodeled dynamics belongs to this type of perturbation. On the other hand, when $\delta(t)>0$ but $\gamma(t)\equiv 0$, the perturbation is \emph{persistent}. External disturbance belongs to this type of perturbation. 

We also remark here that the assumptions on $g(t,x)$ as stated in Assumption~\ref{ass:3}, together with the boundedness assumption of $A(t)$ in Assumption~\ref{ass:1} and the piece-wise continuity assumption of $A(t)$ in Assumption~\ref{ass:2}.2 imply that the right-hand side of \eqref{def:ltv_sys} satisfies the Carath\'eodory's condition for existence and uniqueness of local solutions for each initial pair $t_0,x_0$ \cite[Page 30]{hale_1980}, and therefore our system \eqref{def:ltv_sys} is well-defined under these assumptions.

\subsection{Slowly varying by means of small total variation}\label{subsec:slow_variation}
Just as in the work \cite{Gao2018}, we quantify the slow time-varying nature of the system \eqref{def:ltv_sys} by imposing bounds on the total variation of $A(t)$ over an arbitrary interval $[a,b]$. Recall the definition of total variation in \eqref{def:total_variation}, which involves a supremum over an uncountable set and is difficult to utilize. \revise{Nevertheless, the following Lemma gives a convenient formula for computing the total variation of $A(t)$ over $[a,b]$ when Assumption~\ref{ass:2}.1 to Assumption~\ref{ass:2}.3 hold:
\begin{lem}\label{lem:TV_formula}
Consider a matrix valued function $A(t)$ satisfying 	Assumption~\ref{ass:2}.1 to Assumption~\ref{ass:2}.3. The total variation of $A(t)$ over $[a,b]$ is given by the following expression:
\begin{equation}\label{dA}
	\int_{a}^{b}\Vert dA\Vert =\sum_{i=0}^{p-1}\int_{t_i}^{t_{i+1}}\Vert \dot A(t)\Vert dt+\sum_{i=1}^{p}\Vert  A(t_i)- A(t_i^-)\Vert.
\end{equation}
\end{lem}
The proof of Lemma~\ref{lem:TV_formula} is given in Appendix~\ref{sec:pf:TV}.
\begin{rem}
	As stated by \cite[Proposition 3.8]{Leoni2017}, absolute continuity of $f(t)$ means its weak derivative $\dot f(t)$ is Lebesgue integrable. This result can be easily extended to matrix-valued functions and therefore $\int_{t_i}^{t_{i+1}}\Vert \dot A(t)\Vert dt$ in \eqref{dA} is well-defined. Compared with \cite[Lemma 1]{Gao2018}, although Lemma~\ref{lem:TV_formula} provides the same formula for total variation, because our assumption is strictly weaker than the one used in \cite{Gao2018} (piece-wise absolutely continuous versus piece-wise continuously differentiable and the derivative is Riemann integrable), the proofs of the two results are different.
\end{rem}}

In addition to the consideration of the variation of $A(t)$, because our system is allowed to have unstable instantaneous dynamics, the variation of $\alpha(A(t))$ when it is not sufficiently negative also needs to be taken into account when defining the slow time-varying nature of the system. To this end, we first define the ramp function $f_{\ramp}(s):\R\mapsto\Rp$ by $f_{\ramp}(s)=\max\{s,0\}$. For any $\kappa>0$, Define $\varphi_{\kappa}(A):\R^{n\times n}\to \Rp$ by
\begin{equation}\label{def:varphi}
    \varphi_{\kappa}(A):=f_{\ramp}(\alpha(A)+\kappa).
\end{equation}
By this definition, $\varphi_{\kappa}(A)=0$ if $\alpha(A)\leq -\kappa$. We then study the total variation of $\varphi_{\kappa}(A)$ over an arbitrary interval $[a,b]$, denoted by $\int_{a}^{b}|d\varphi_{\kappa}(A)|$. 
It is not difficult to see that for any $t,s\in[a,b]$, $|\varphi_{\kappa}(A(t))-\varphi_{\kappa}(A(s))|\leq |\alpha(A(t))-\alpha(A(s))|$. Thus by Assumption~\ref{ass:2}.4 and Lemma~\ref{lem:Lipschitz} in Appendix~\ref{sec:lem:ac}, we conclude that $\varphi_{\kappa}(A(t))$ is also piece-wise absolutely continuous. Thus by a similar proof of Lemma~\ref{lem:TV_formula}, we have
\begin{multline}\label{d_phi_A}
        \int_{a}^{b}|d\varphi_{\kappa}(A)| =\sum_{i=0}^{p-1}\int_{t_i}^{t_{i+1}}|\dot\varphi_{\kappa}(A(t))| dt\\
        +\sum_{i=1}^{p}| \varphi_{\kappa}(A(t_i))- \varphi_{\kappa}(A(t_i^-))|.
\end{multline}

We can also study the slowly time-varying nature of \eqref{def:ltv_sys} by considering the combined total variations of $A(t)$ and $\varphi_{\kappa}(A(t))$ instead of studying them seperately. To do this, define the matrix trajectory $\tilde A(t):\Rp\mapsto\R^{n\times n}$ by 
\begin{equation}\label{def:tilde_A}
    \tilde A(t):=A(t)-\varphi_{\kappa}(A(t))I.
\end{equation} 
We have the following conclusion:
\begin{prop}\label{prop:0}
Consider a matrix trajectory  $A(t):\Rp\mapsto \R^{n\times n}$. For some $\kappa>0$, let $\varphi_{\kappa}$ be defined by \eqref{def:varphi} and $\tilde A(t)$ be defined by \eqref{def:tilde_A}. Under Assumption~\ref{ass:2}, the total variation of $\tilde A(t)$ over the interval $[a,b]$ is well-defined and satisfies
\begin{equation}\label{relation_d_tilde_A_dA_d_phi_A}
    \int_{a}^{b}\Vert d\tilde A\Vert\leq\int_{a}^{b}\Vert d A\Vert+ \int_{a}^{b}|d\varphi_{\kappa}(A)|,
\end{equation}
where $\int_{a}^{b}\Vert d A\Vert, \int_{a}^{b}|d\varphi_{\kappa}(A)|$ are given by \eqref{dA}, \eqref{d_phi_A} respectively.
\end{prop}
\begin{pf}
By the construction \eqref{def:tilde_A}, $\tilde A(t)$ satisfies the same regularity assumptions on $A(t)$ (C\`adl\`ag, finite discontinuities and piece-wise absolute continuity) and hence
\begin{equation}\label{d_tilde_A}
    \int_{a}^{b}\Vert d\tilde A\Vert =\sum_{i=0}^{p-1}\int_{t_i}^{t_{i+1}}\Vert\dot{\tilde A}(s)\Vert ds+\sum_{i=1}^{p}\Vert  \tilde A(t_i)- \tilde A(t_i^-)\Vert.
\end{equation}
Meanwhile, for almost all $t\not\in\cD$,
\[
\Vert\dot{\tilde A}(t)\Vert=\Vert\dot A(t)-\dot\varphi_{\kappa}(A(s))I\Vert\\
\leq\Vert\dot A(t)\Vert+|\dot\varphi_{\kappa}(A(t))|   
\]
and for all $t\in\cD$,
\begin{align*}
\Vert  \tilde A(t)&- \tilde A(t^-)\Vert\\
&=\Vert(A(t)-A(t^-))-(\varphi_{\kappa}(A(t))- \varphi_{\kappa}(A(t^-)))I\Vert\\
&\leq    \Vert(A(t)-A(t^-))\Vert+|\varphi_{\kappa}(A(t))- \varphi_{\kappa}(A(t^-))|.
\end{align*}
Plug these upper bounds into \eqref{d_tilde_A} and appeal to the expressions \eqref{dA}, \eqref{d_phi_A}, the inequality \eqref{relation_d_tilde_A_dA_d_phi_A} is hence shown.
\end{pf}

\subsection{Other necessary technical results}
Note that so far we have not invoked Assumption~\ref{ass:1}. In fact under Assumption~\ref{ass:1} we have the following result: 
\begin{lem}\label{lem:from_khalil}
Consider a matrix trajectory  $A(t):\Rp\mapsto \R^{n\times n}$. For some $\kappa>0$, let $\varphi_{\kappa}$ be defined by \eqref{def:varphi} and $\tilde A(t)$ be defined by \eqref{def:tilde_A}. Under Assumption~\ref{ass:1}, for any $\beta\in(0,\kappa)$, there exists $c=c(L,\kappa,\beta)>0$ such that
\begin{equation}\label{from_khalil_part_1}
    \Vert e^{s\tilde A(t)}\Vert \leq ce^{-\beta s}\quad\forall s,t\in\Rp.
\end{equation}
Meanwhile, the Lyapunov equation
\begin{equation}\label{Lyapunov_equation}
    \tilde A(t)^\top P+P\tilde A(t)+I=0
\end{equation}
has a unique solution $P(t)$ for each $t\in\Rp$ and
\begin{subequations}\label{sandwich_P}
\begin{align}
c_1&\leq \Vert P(t)\Vert \leq c_2\quad\forall t\in\Rp,\\
 c_1|z|^2&\leq z^\top P(t) z\leq c_2|z|^2\quad\forall z\in\R^n,t\in\Rp
\end{align}
\end{subequations}
with 
\begin{equation}\label{def:c_1_c_2}
    c_1:=\frac{1}{2(L+f_{\ramp}(L+\kappa))},c_2:=\frac{c^2}{2\beta}.
\end{equation}
\revise{Moreover, if $\tilde A(t)$ is absolutely continuous over $[a,b]$, then $P(t)$ is absolutely continuous over $[a,b]$ as well} and
\begin{equation}\label{bound_derivative_P}
    \Vert\dot P(t)\Vert\leq 2c_2^2\Vert\dot{\tilde A}(t)\Vert
\end{equation}
for almost all $t\in[a,b]$.
\end{lem}
The proof of Lemma~\ref{lem:from_khalil} is provided in Appendix~\ref{sec:pf:P}.
We also need the following result which bounds the difference in $P(a), P(b)$ in terms of $\tilde A(a),\tilde A(b)$:

\begin{lem}[{\cite[Proposition 1]{Gao2018}}]\label{lem:from_Xiaobin}
Consider a matrix trajectory  $A(t):\Rp\mapsto \R^{n\times n}$. For some $\kappa>0$, let $\varphi_{\kappa}$ be defined by \eqref{def:varphi} and $\tilde A(t)$ be defined by \eqref{def:tilde_A}. Assume Assumption~\ref{ass:1} holds on $A(t)$ and consider the function $V(t,x):=x^\top P(t) x$ for each $(t,x)\in\Rp\times\R^n$, where $P(t)$ is the solution to \eqref{Lyapunov_equation}. Then, for any $a,b\geq 0$,
\begin{align}
    &\Vert P(b)-P(a)\Vert\leq 2c_2^2\Vert \tilde A(b)-\tilde A(a)\Vert,\label{bounded_P_change}\\
    &V(b,x)\leq e^{2c_2^2c_1^{-1}\Vert \tilde A(b)-\tilde A(a)\Vert}V(a,x)\quad\forall x\in \R^n.\label{bound_jump}
    \end{align}
where $c_1,c_2$ come from Lemma~\ref{lem:from_khalil}.
\end{lem}

\section{Stability of slowly time-varying system}\label{sec:thm}
We now state our main result:
\begin{thm}\label{thm:main}
Consider an LTV system with perturbation \eqref{def:ltv_sys} with Assumption~\ref{ass:1}, Assumption~\ref{ass:2} and Assumption~\ref{ass:3} satisfied. Let $\kappa>0$. Then if there exist $\lambda<\frac{c_1}{2c_2},\varrho>0$, such that for all $b> a\geq 0$,
\begin{multline}\label{GE-DSS_mixed_condition}
    c_1\int_{a}^{b}\varphi_{\kappa}(A(\tau))d\tau+c_2\int_{a}^{b}\gamma(\tau)d\tau+c_2^2\Big(\int_{a}^{b}\Vert d \tilde A\Vert\Big)\\
    \leq \lambda(b-a)+\varrho
\end{multline}

where $c_1,c_2$ are defined via \eqref{def:c_1_c_2} in Lemma~\ref{lem:from_khalil}, then there exists $k_1,k_2,k_3>0$ such that
\begin{equation}\label{ISS}
|x(t;t_0,x_0)|\leq k_1 e^{-k_2(t-t_0)}|x_0|+k_3\max_{\tau\in[t_0,t]}\delta(\tau).
\end{equation}
\end{thm}

\subsection{Discussion of Theorem~\ref{thm:main}}

We give some insights of Theorem~\ref{thm:main} before we proceed to its proof. 

We start with the discussion on the estimate \eqref{ISS} first. This result actually implies that the system \eqref{def:ltv_sys} is uniformly \emph{input-to-state stable} (ISS) with respect to the origin (see the definition of ISS in \cite{Sontag1996}), where the ``input" is the persistent part of the perturbation $\delta(t)$. ISS also implies that the system \eqref{def:ltv_sys} has the ``convergent input convergent state" property, meaning that if $\lim_{t\to\infty}\delta(t)=0$, then the solutions of \eqref{def:ltv_sys} will converge to the origin. When the perturbation is vanishing such that $\delta(t)\equiv 0$, \eqref{ISS} also shows that the system \eqref{def:ltv_sys} is uniformly globally exponentially stable.

We then turn to the condition \eqref{GE-DSS_mixed_condition}. The three terms in \eqref{GE-DSS_mixed_condition} on the left-hand side are essentially the total effect of unstable $A(t)$, the total estimate of perturbation-to-state ratio and the total variation of $\tilde A$ over the interval $[a,b]$. We discuss some special cases here and compare them with the known results from the literature.
\begin{itemize}
\item Assume that $A(t)$ is always Hurwtiz and $\alpha(A(t))\leq -\kappa^*$ for some $\kappa^*>0$ and all $t\geq t_0$. In this case we can pick $\kappa=\kappa^*$, which implies that $\varphi_{\kappa}(A(t))\equiv 0$. If in addition we assume that the system is unperturbed, i.e., $\gamma(t)\equiv 0$, then \eqref{GE-DSS_mixed_condition} reduces to 
\[
\Big(\int_{a}^{b}\Vert d \tilde A\Vert\Big)  \leq \frac{\lambda}{c_2^2}(b-a)+\frac{\varrho}{c_2^2}.
\]
Note that the upper-bound on $\lambda$ stated in Theorem~\ref{thm:main} implies that $\mu:=\frac{\lambda}{c_2^2}\leq\frac{c_1}{2c_2^3}$. Thus we recover exactly the same criteria as in \cite[Theorem 3]{Gao2018} for testing global exponential stability of LTV systems with bounded total variation. Furthermore, if $A(t)$ is continuously differentiable, then this result becomes the same as \cite[Theorem 3.4.11]{ioannou_sun_1996}. 

\item Now we assume $A(t)=A$ is a constant Hurwitz matrix. By picking $\kappa=\alpha(A)$, we have $\varphi_{\kappa}(A(t))\equiv 0$ and $\tilde A=A$ so $\int_{a}^{b}\Vert d \tilde A\Vert=0$. Moreover, the time-invariant Lyapunov function $V(x):=x^\top P x$ has the property that 
\[
\dot V(x)\leq -c_3|x|^2, \quad |\nabla V(x)|\leq c_4|x|
\]
with the parameters $c_3=1,c_4=2c_2$. 
In the presence of perturbation, the condition \eqref{GE-DSS_mixed_condition} reduces to
\[
\int_{a}^{b}\gamma(\tau)d\tau\leq \frac{\lambda}{c_2}(b-a)+\frac{\varrho}{c_2}.
\]
Moreover, the upper-bound on $\lambda$ implies $\epsilon:=\frac{\lambda}{c_2}\leq \frac{c_1}{2c_2^2}=\frac{c_1c_3}{c_2c_4}$. This is exactly the same results as \cite[Lemma 9.4 and Corollary 9.1]{Khalil2002} for showing global exponential stability with respect to a neighborhood of the origin for a perturbed system.

\item Lastly, consider a switched system with linear subsystems
\[
\dot x(t)=A_{\sigma(t)}x(t)
\]
where $\sigma(t):\Rp\mapsto \cP:=\{1,\cdots,p\}$ is a C\`adl\`ag piece-wise constant function. We assume that there exist $\alpha_s,\alpha_u>0$ and a partition $\cP=\cP_s\cup\cP_u$ such that $\alpha(A_i)\leq -\alpha_s$ for all $i\in\cP_s$ and $\alpha(A_i)\leq \alpha_u$ for all $i\in\cP_u$. In other words, not all subsystems are assumed to be stable. In this case we pick $\kappa=\alpha_s$. We further assume that there exists $\Delta A>0$ such that $\Vert\tilde A_i-\tilde A_j\Vert\leq \Delta A$ for all $i,j\in\cP$, where $\tilde A_i:=A_i-\varphi_{\kappa}(A_i)I$. It can be concluded that
\[
\int_{a}^{b}\varphi_\kappa(A_{\sigma(\tau)})d\tau\leq(\alpha_s+\alpha_u)\int_{a}^{b}{\mathbf 1}(\tau)d\tau,
\]
where ${\mathbf 1}(\tau)$ is the indicator function for $\sigma(\tau)\in\cP_u$, and
\[
\int_{a}^{b}\Vert d \tilde A\Vert\leq \Delta A\#(\cD\cap(a,b)),
\]
where $\#(\cdot)$ denotes the cardinality of a set. \revise{Now suppose the switching signal $\sigma(t)$ satisfies some average dwell-time condition \cite{Hespanha1999} and average activation time condition \cite{Mueller2012}; that is, there exist $\tau_a,\eta,N_0,T_0>0$ such that
\begin{align*}
	&\#(\cD\cap(a,b))\leq \frac{1}{\tau_a}(b-a)+N_0,\\
	&\int_{a}^{b}{\mathbf 1}(\tau)d\tau\leq \eta(b-a)+T_0
\end{align*}
for any $b>a\geq 0$. If
\begin{equation}\label{specified_AAT_ADT}
	\lambda:=c_1(\alpha_s+\alpha_u)\eta+\frac{c_2^2\Delta A}{\tau_a}<\frac{c_1}{2c_2},
\end{equation}
then it is not difficult to verify that \eqref{GE-DSS_mixed_condition} will hold with $\varrho:=c_1(\alpha_s+\alpha_u)T_0+c_2^2\Delta AN_0$. 
On the other hand, pick the Lyapunov functions $V_i(x)=x^\top P_i x$, where $P_i$ is the solution to $P_i\tilde A_i+\tilde A_i^\top P_i+I=0$ for each $i\in\cP$. Then we have
\begin{align*}
	\dot V_i(x)&=-|x|^2\leq -\frac{1}{c_2}V_i(x)&\quad\forall i\in\cP_s,\\
	\dot V_i(x)&\leq2(\alpha_s+\alpha_u)x^\top P_ix-|x|^2\\
	&\leq\left(2(\alpha_s+\alpha_u)-\frac{1}{c_2}\right)V_i(x)&\quad\forall i\in\cP_u.
\end{align*} 
Meanwhile, it follows from \eqref{bounded_P_change} in Lemma~\ref{lem:from_Xiaobin} that $\Vert P_i-P_j\Vert\leq 2c_2^2\Vert\tilde A_i-\tilde A_j\Vert\leq 2c_2^2\Delta A$ so
\[
P_i=P_i-P_j+P_j\leq \Vert P_i-P_j\Vert I+P_j\leq \left(\frac{2c_2^2\Delta A}{c_1}+1\right)P_j.
\]
As a result, the assumptions in \cite[Theorem 2]{Mueller2012} are satisfied with $\lambda_s:=\frac{1}{c_2},\lambda_u:=2(\alpha_s+\alpha_u)-\frac{1}{c_2}$ and $\mu:=\frac{2c_2^2\Delta A}{c_1}+1$. For an unforced system, that theorem (see also \cite[Page 8]{SL-AT-DL:21}) essentially implies that if
\begin{equation}\label{eqn:from_Mueller}
\left(1+\frac{\lambda_u}{\lambda_s}\right)\eta+\frac{\ln\mu}{\lambda_s\tau_a}<1,
\end{equation}
then the system is globally asymptotically (in fact exponentially) stable. By plugging the definitions of $\lambda_s,\lambda_u,\mu$ into \eqref{eqn:from_Mueller} and multiplying both sides by $\frac{c_1}{2c_2}$, we have
\begin{equation*}
	c_1(\alpha_s+\alpha_u)\eta+\frac{c_1\ln\left(\frac{2c_2^2\Delta A}{c_1}+1\right)}{2\tau_a}<\frac{c_1}{2c_2},
\end{equation*}
which is a necessary condition for \eqref{specified_AAT_ADT} since $\frac{2c_2^2\Delta A}{c_1}\geq\ln\left(\frac{2c_2^2\Delta A}{c_1}+1\right)$. Thus Theorem~\ref{thm:main} is related to the known results for switched system with unstable modes in the literature, in the sense that under the same condition \eqref{specified_AAT_ADT} on average dwell-time and average activation time, either Theorem~\ref{thm:main} or other approaches from the literature can be used to show stability of the system.}
\end{itemize}


As illustrated by the aforementioned comparisons, our result is a generalization of the known results in the literature. 
We also point out that when the matrix $\tilde A$ is not directly accessible, thanks to Proposition~\ref{prop:0}, a sufficient condition for \eqref{GE-DSS_mixed_condition} to hold is the following:
\begin{multline*}
    c_1\int_{a}^{b}\varphi_{\kappa}(A(\tau))d\tau+c_2\int_{a}^{b}\gamma(\tau)d\tau\\+c_2^2\Big(\int_{a}^{b}\Vert d A\Vert+\int_{a}^{b}|d\varphi_{\kappa}(A)|\Big)    \leq \lambda(b-a)+\varrho,
\end{multline*}
where one only needs to evaluate $A(t)$, $\varphi_k(A(t))$ and $\gamma(t)$.

\subsection{Proof of Theorem~\ref{thm:main}}

The proof essentially contains two steps. In the first step we will define a function $\xi(t):\Rp\mapsto\R$ which traces the change of the left-hand side of \eqref{GE-DSS_mixed_condition} and show that it is piece-wise absolutely continuous and always in a bounded set. In the second step we will use $\xi(t)$ to construct a time-varying Lyapunov function which is monotonically decreasing when $\delta(t)\equiv 0$, and hence use it to show the desired property \eqref{ISS}.

We start the first step of the proof by defining
\begin{equation}\label{def:xi}
    \xi(t):=\inf_{s\in[0,t]}\chi(s)-\chi(t)+\varrho,
\end{equation}
where $\chi(t):=c_1\int_0^t\varphi_{\kappa}(A(\tau))d\tau+c_2\int_0^t\gamma(\tau)d\tau+c_2^2\int_0^t\Vert d \tilde A\Vert-\lambda t$. By its definition, $\xi(t)\leq \varrho$ for all $t\in\Rp$. In addition, it follows from \eqref{GE-DSS_mixed_condition} that for any $s\in[0,t]$,
\begin{multline*}
    \chi(s)-\chi(t)+\varrho=-c_1\int_s^t\varphi_{\kappa}(A(\tau))d\tau-c_2\int_s^t\gamma(\tau)d\tau\\
    -c_2^2\int_s^t\Vert d \tilde A\Vert+\lambda (t-s)+\varrho\geq 0.
\end{multline*}
Therefore
\begin{equation}\label{sandwich_xi}
    0\leq\xi(t)\leq \varrho.
\end{equation}
\revise{Under Assumption~\ref{ass:2}, $\tilde A(t)$ is absolutely continuous over each interval $[t_{i},t_{i+1})$ where it is continuous. Thus for $t\in[t_{i},t_{i+1})$, $\chi(t)-\chi(t_i)$ is the sum of integration of Lebesgue integrable functions. Therefore, $\chi(t)$ is absolutely continuous over $[t_{i},t_{i+1})$.} 
Moreover, since absolute continuity is preserved for $\inf_{s\in[0,t]}\chi(s)$, $\xi(t)$ is also absolutely continuous over $[t_{i},t_{i+1})$.
\revise{Note that since $\inf_{s\in[0,t]}\chi(s)$ is non-increasing, it follows from \eqref{def:xi} that for any $r\geq t$,
\begin{align}
	\xi(r)-\xi(t)&=\big(\inf_{s\in[0,r]}\chi(s)-\inf_{s\in[0,t]}\chi(s)\big)-(\chi(r)-\chi(t))\nonumber\\
	&\leq -(\chi(r)-\chi(t)),\label{diff_xi}
\end{align}
Let both $r,t\in(t_i,t_{i+1})$. Divide \eqref{diff_xi} by $r-t$ and take the limit as $r-t\to 0$, we conclude that}
\begin{multline}
\dot\xi(t)\leq -\dot\chi(t)\\
=-c_1\varphi_{\kappa}(A(t))-c_2\gamma(t)-c_2^2\Vert\dot{\tilde A}(t)\Vert+\lambda\ \forall\mbox{ a.a. }t\not\in\cD,\label{xi_flow}
\end{multline}
where recall $\cD$ defined in \eqref{def:D} is the set of discontinuities. \revise{On the other hand, let $r=t_i$ and take the limit as $t\to t_i^-$, we have $\chi(t_i^-)=\lim_{t\to t_i^-}\chi(t)=\chi(t_i)-c_2^2\Vert\tilde A(t_i)-A(t_i^-) \Vert$
so}
\begin{equation}
\xi(t)-\xi(t^-)\leq-c_2^2\Vert  \tilde A(t)- \tilde A(t^-)\Vert\ \forall t\in\cD.\label{xi_jump}
\end{equation}
Now we proceed to the second step of the proof. Define the function $U(t):\Rp\mapsto\Rp$ with $U(t):=e^{\frac{2\xi(t)}{c_1}}$, and two functions $V(t,x),W(t,x):\Rp\times \R^{n}\mapsto\Rp$ such that $V(t,x):=x^\top P(t) x$ where $P(t)$ is the solution to the Lyapunov equation \eqref{Lyapunov_equation}, and $W(t,x):=U(t)V(t,x)$. We will show that $W(t,x)$ is the desired time-varying Lyapunov function.
It follows from \eqref{sandwich_P} and \eqref{sandwich_xi} that 
\begin{equation}\label{sandwich_W}
c_1|x|^2\leq W(t,x)\leq c_2e^{\frac{2\varrho}{c_1}}|x|^2\quad\forall (t,x)\in \Rp\times \R^{n}.
\end{equation}
\revise{Recall in Lemma~\ref{lem:from_khalil},  $P(t)$ is shown to be piece-wise absolutely continuous. Meanwhile it is already discussed earlier that $\xi(t)$ is also piece-wise absolutely continuous and bounded. Therefore $W(t,x(t))$ is piece-wise absolutely continuous.} We investigate the weak time derivative of $W(t,x(t))$ for $t\not\in \cD$ and the jump of $W(t,x)$ for $t\in \cD$ separately. To this end, it follows from \eqref{xi_flow} that
\begin{multline*}
\dot U(t)=\frac{2}{c_1}U(t)\dot\xi(t)\\
\leq \big(-2\varphi_{\kappa}(A(t))-\frac{2c_2}{c_1}\gamma(t)-\frac{2c_2^2}{c_1}\Vert\dot{\tilde A}(t)\Vert+\frac{2\lambda}{c_1}\big)U(t).    
\end{multline*}
Next we estimate the time derivative of $V(t,x(t))$. For simplification we omit the argument $t$ when there is no ambiguity. Note that since $\lambda<\frac{c_1}{2c_2}$, there exists $\epsilon\in(0,\frac{c_1}{c_2}-2\lambda)$. 
It then follows from \eqref{sandwich_P},  \eqref{bound_derivative_P} and Assumption~\ref{ass:3} that
{\small\begin{align*}
    \dot V&(t,x(t))=\big(Ax+g(t,x)\big)^\top Px+x^\top P \big(Ax+g(t,x)\big)+x^\top \dot Px\\
    &=x^\top \left(A^\top P+PA\right)x+2x^\top Pg(t,x)+x^\top \dot Px\\
    &\leq x^\top \big(-I+2\varphi_{\kappa}(A)P\big)x+2|x|\Vert P\Vert(\gamma|x|+\delta)+2c_2^2\Big\Vert\dot{\tilde A}\Big\Vert |x|^2\\
    &= 2\varphi_{\kappa}(A)x^\top Px+2|x|\Vert P\Vert\delta+\left(2c_2^2\Big\Vert\dot{\tilde A}\Big\Vert-1+2\Vert P\Vert\gamma\right)|x|^2\\
        &\leq 2\varphi_{\kappa}(A)x^\top Px+\frac{c_2^2}{\epsilon}\delta^2+\left(2c_2^2\Big\Vert\dot{\tilde A}\Big\Vert-1+2c_2\gamma+\epsilon \right)|x|^2\\
&\leq \left(2\varphi_{\kappa}(A)+\frac{2c_2^2}{c_1}\Big\Vert\dot{\tilde A}\Big\Vert-\frac{1}{c_2}+\frac{2c_2}{c_1}\gamma+\frac{\epsilon }{c_1}\right)V+\frac{c_2^2}{\epsilon}\delta^2.
\end{align*}}
\revise{where $2|x|\Vert P\Vert\delta\leq \epsilon|x|^2+\frac{\Vert P\Vert^2\delta^2}{\epsilon}\leq \epsilon|x|^2+\frac{c_2^2}{\epsilon}\delta^2$ is used for the second inequality above.}
Using product rule and plug the bounds on $\dot U$ and $\dot V$ in,
\begin{align*}
    \dot W&(t,x(t))=\dot U(t)V(t,x(t))+U(t)\dot V(t,x(t)\\
    &\leq \big(\frac{2\lambda+\epsilon}{c_1}-\frac{1}{c_2}\big)W(t,x(t))+ \frac{c_2^2}{\epsilon}U(t)\delta(t)^2\\
    &\leq \big(\frac{2\lambda+\epsilon}{c_1}-\frac{1}{c_2}\big)W(t,x(t))+ \frac{c_2^2}{\epsilon}e^{\frac{2\varrho}{c_1}}\delta(t)^2.
\end{align*}
In other words, we conclude that
\begin{equation}\label{W_flow}
    \dot W(t,x(t))\leq -a_0W(t,x(t))+ b_0|\delta(t)|^2\ \forall\mbox{a.a. } t\not\in\cD
\end{equation}
where $a_0:=\frac{1}{c_2}-\frac{2\lambda+\epsilon}{c_1}>0$ and $b_0:=\frac{c_2^2}{\epsilon}e^{\frac{2\varrho}{c_1}}$. 

Now for all $t\in\cD$, it follows from \eqref{xi_jump} that
\[
U(t)\leq e^{-2c_2^2c_1^{-1}\Vert  \tilde A(t)- \tilde A(t^-)\Vert}U(t^-).
\]
Meanwhile, since the solution is continuous, $x(t)=x(t^-)$ and it follows from \eqref{bound_jump} in Lemma~\ref{lem:from_Xiaobin} that
\[
V(t,x(t))\leq e^{2c_2^2c_1^{-1}\Vert \tilde A(t)-\tilde A(t^-)\Vert}V(t^-,x(t^-)).
\]
Therefore
\begin{multline}\label{W_jump}
W(t,x(t))=U(t)V(t,x(t))\leq U(t^-)V(t^-,x(t^-))\\
=W(t^-,x(t^-))\quad\forall t\in\cD.
\end{multline}

From the estimates \eqref{W_flow}, \eqref{W_jump} and comparison principle, we conclude that for any $t\geq t_0\geq 0$,
\begin{align*}
W(t,x(t))&\leq e^{-a_0(t-t_0)}W(t_0,x_0)+b_0\int_{t_0}^te^{-a_0(t-\tau)}\delta(\tau)^2d\tau\\
&\leq e^{-a_0(t-t_0)}W(t_0,x_0)+\frac{b_0}{a_0}\max_{\tau\in[t_0,t]}\delta(\tau)^2.
\end{align*}
Finally, it follows from \eqref{sandwich_W} and the inequality $\sqrt{a_1+a_2}\leq \sqrt{a_1}+\sqrt{a_2}$ that
\begin{align*}
|x(t)|&\leq \sqrt{\frac{W(t,x(t))}{c_1}}\\
&\leq e^{-\frac{a_0}{2}(t-t_0)}\sqrt{\frac{W(t_0,x_0)}{c_1}}+\sqrt{\frac{b_0}{a_0}}\max_{\tau\in[t_0,t]}\delta(\tau)\\
&\leq \sqrt{\frac{c_2}{c_1}}e^{\frac{\varrho}{c_1}-\frac{a_0}{2}(t-t_0)}|x_0|+\frac{b_0}{a_0}\max_{\tau\in[t_0,t]}\delta(\tau)
\end{align*}
We thus achieve \eqref{ISS} with $k_1:=\sqrt{\frac{c_2}{c_1}}e^{\frac{\varrho}{c_1}},k_2:=\frac{a_0}{2}$ and $k_3:=\sqrt{\frac{b_0}{a_0}}$. This concludes the proof of Theorem~\ref{thm:main}.

\begin{rem}\label{rem:final}
\revise{Observe that in Lemma~\ref{lem:from_Xiaobin}, Assumption~\ref{ass:1} guarantees the existence of $c_1,c_2$ such that \eqref{sandwich_P} and \eqref{bound_derivative_P} hold. These inequalities are used in the proof of Theorem~\ref{thm:main}. In other words, we do not necessarily require $\Vert A(t)\Vert$ to be bounded. Instead, if we replace Assumption~\ref{ass:1} with the assumptions that the spectrum of $P(t)$ is uniformly bounded in an interval, and $\Vert\dot P(t)\Vert$ is relatively uniformly bounded with respect to $\Vert\dot{\tilde A}(t)\Vert$, then we could conclude the same result as in Theorem~\ref{thm:main}.}
\end{rem}

\section{A numerical example}\label{sec:example}
Consider a 2-dimensional periodic system with period $2\pi$. \revise{For any $k\in\mathbb Z$ and $t\in [0,2\pi)$, the dynamics is given by
\begin{equation}\label{sys:example}
\dot x(2k\pi+t)=\begin{pmatrix}
\lambda(t)&0.1\cos(t)+1\\0.1\cos(t)-1&\lambda(t)
\end{pmatrix}x(2k\pi+t).
\end{equation}
where $\lambda(t):=1.1\cos(\frac{t}{2})+0.1\sin(t)-1$}. We can re-write \eqref{sys:example} in the form of \eqref{def:ltv_sys}, with
\begin{align*}
A(t)&:=\begin{pmatrix}
1.1\cos(\frac{t}{2})-1&1\\-1&1.1\cos(\frac{t}{2})-1
\end{pmatrix},\\
 g(t,x)&:=0.1\begin{pmatrix}
\sin(t)&\cos(t)\\\cos(t)&\sin(t)
\end{pmatrix}x.
\end{align*}
Note that $A(t)$ is not continuous, since $A(2\pi^-)=\begin{pmatrix}
-2.1&1\\-1&-2.1
\end{pmatrix}$ but $A(2\pi)=A(0)=\begin{pmatrix}
0.1&1\\-1&0.1
\end{pmatrix}$. Moreover, $\alpha(A(t))=1.1\cos(\frac{t}{2})-1$. Since $\alpha(A(0))=0.1$, the system \eqref{sys:example} has unstable instantaneous dynamics.

To apply Theorem~\ref{thm:main}, we take $\kappa=1$. Since the system is periodic, we will only investigate \eqref{GE-DSS_mixed_condition} over one period; i.e., $a=0$ and $b=2\pi$. It can be found that 
\[
\varphi_{\kappa}(A(t))=\begin{cases}
1.1\cos(\frac{t}{2})&\text{ for }t\in[0,\pi),\\
0&\text{ for }t\in[\pi,2\pi).
\end{cases}
\]
Therefore $\int_0^{2\pi}\varphi_{\kappa}(A(t))dt=2.2$. \revise{Meanwhile, 
\[
\tilde A(t)=\begin{cases}
	\begin{pmatrix}
		-1&1\\-1&-1
	\end{pmatrix}&\mbox{ if }t\in[0,\pi),\\
	A(t)&\mbox{ if }t\in[\pi,2\pi).
\end{cases}
\]
In order to find $c_1,c_2$ which will be used in Theorem~\ref{thm:main}, we recall Remark~\ref{rem:final}. Thus instead of setting \eqref{def:c_1_c_2} as in Lemma~\ref{from_khalil_part_1}, we solve \eqref{Lyapunov_equation} for $P(t)$:
\[
P(t)=\begin{cases}
		\frac{1}{2}I&\mbox{ if }t\in[0,\pi),\\
	\frac{1}{2}(1-1.1\cos(\frac{t}{2}))^{-1}I&\mbox{ if }t\in[\pi,2\pi).
\end{cases}
\]
Therefore, 
\begin{align*}
	&\frac{1}{4.2}\leq\Vert P(t)\Vert\leq\frac{1}{2},\\
	&\Vert\dot P(t)\Vert=\frac{1}{2}\left(1-1.1\cos(\frac{t}{2})\right)^{-2}\Vert\dot{\tilde A}(t)\Vert\leq \frac{1}{2}\Vert\dot{\tilde A}(t)\Vert
\end{align*}
so \eqref{sandwich_P} and \eqref{bound_derivative_P} hold with $c_1=0.2381, c_2=0.5$.} Meanwhile,
\[
\int_0^{2\pi}\Vert d\tilde A\Vert=2\left\Vert\begin{pmatrix}
-1&1\\-1&-1
\end{pmatrix}-\begin{pmatrix}
-2.1&1\\-1&-2.1
\end{pmatrix}\right\Vert=2.2.
\]
In addition, $|g(t,x)|\leq 0.1\left\Vert\begin{pmatrix}
\sin(t)&\cos(t)\\\cos(t)&\sin(t)
\end{pmatrix}\right\Vert|x|= 0.1(|\cos(t)|+|\sin(t)|)|x|$. Hence the inequality \eqref{bound_on_g} holds with $\gamma(t)=0.1(|\cos(t)|+|\sin(t)|)$ and $\delta(t)\equiv 0$. We thus have $\int_0^{2\pi}\gamma(t)dt=0.8$. Consequently, the left-hand side of \eqref{GE-DSS_mixed_condition} gives $1.4738$, while by picking $\lambda=0.238<\frac{c_1}{2c_2}$ and $\varrho=0$, the right-hand side of \eqref{GE-DSS_mixed_condition} gives $1.4954$. Finally we remark here that even if $[a,b]$ is not a multiple of period $2\pi$, the discrepancies in the integration can always be bounded by \revise{$\varrho=1.4738$, which is the value of the left-hand side of \eqref{GE-DSS_mixed_condition} over one period.} Therefore \eqref{GE-DSS_mixed_condition} always holds. Because $\delta(t)\equiv 0$, we conclude from Theorem~\ref{thm:main} that the system \eqref{sys:example} is uniformly globally exponentially stable.

\section{Discussion and conclusion}\label{sec:conclusion}
In this work the stability of perturbed LTV systems is studied. We considered different challenging features for the problem in this work, including the assumption that the system matrix is piece-wise absolutely continuous, the assumption that the instantaneous dynamics can be unstable and the assumption that the perturbation might be persistent. With the help of the characterization of bounded total variation of the matrix trajectory $A(t)$, and the construction of a special Lyapunov function which does not increase when $A(t)$ jumps, we managed to propose unified criteria based on the total assessment of all the three aspects and show that when the criteria are met, the neighborhood of the origin, whose size depends on the magnitude of the persistent perturbation, is uniformly globally exponentially stable for the system.

Through the numerical example studied in this work, we realized that while theoretically our result is elegant, it might have some limitations in application. The condition \eqref{GE-DSS_mixed_condition} proposed in Theorem~\ref{thm:main} can be conservative, because the parameters $c_1,c_2$ might be overestimated. Since $c_1,c_2$ depend on the matrix trajectory $P(t)$ as seen in \eqref{sandwich_P}, we can alternatively consider time-varying parameters instead of constants in order to give tighter estimates. We can also consider better choices of $P(t)$ in the future work, in which direction the recent work \cite{Shenyu2022} may give an idea how to ``smartly" choose $P(t)$ by optimization. 
\revise{On the other hand, it is seen that since the total variation only increases with respect to time, when compared with a time-invariant system, the time-varying nature of the system will only bring negative effect on the stability criteria. Therefore, our result is not suitable for the study of time-varying systems where the variations in fact benefit the stabilization. If instead of using integration of induced norm (which is always non-negative) in the characterization of variations, other measures, such as integration of matrix measure (which can be negative), are used here, then it might be possible to conclude stability results where variations are beneficial. This will be another direction of future work.}

\revise{{\bf Appendix}
\appendix
\section{A useful lemma for absolute continuity}\label{sec:lem:ac}
\begin{lem}\label{lem:Lipschitz}
Consider two vector-valued functions $f(t):[a,b]\mapsto\R^n, g(t):[a,b]\mapsto\R^m$ and suppose $f(t)$ is absolutely continuous on $[a,b]$. If there exists $L>0$ such that 
\begin{equation}\label{Lipschitz}
	|g(c)-g(d)|\leq L|f(c)-f(d)|\quad\forall c,d\in[a,b],
\end{equation}
then $g(t)$ is absolutely continuous on $[a,b]$ as well. Moreover, it holds that $|\dot g(t)|\leq |\dot f(t)|$ for almost all $t\in[a,b]$.
\end{lem}
\begin{pf}
Let $\epsilon>0$ be arbitrary. Define $\epsilon_0:=\frac{\epsilon}{L}$. Since $f(t)$ is absolutely continuous on $[a,b]$, there exists $\delta>0$ such that for any finite sequence of pairwise disjoint sub-intervals $\{(t_k,r_k)\}_k$ of $[a,b]$ satisfying $\sum_k(r_k-t_k)\leq \delta$, one has  $\sum_k|f(r_k)-f(t_k)|\leq \epsilon_0$. It follows from \eqref{Lipschitz} that
\[
\sum_k|g(r_k)-g(t_k)|\leq \sum_kL|f(r_k)-f(t_k)|\leq L\epsilon_0=\epsilon,
\]
which shows that $g(t)$ is absolutely continuous. denote $t:=d,\delta:=c-d$ and divide both sides of \eqref{Lipschitz} by $|\delta|$, we have
\[
\left|\frac{g(t+\delta)-g(t)}{\delta}\right|\leq L\left|\frac{f(t+\delta)-f(t)}{\delta}\right|,
\]
which, by taking the limit as $\delta\to 0$, leads to the conclusion that $|\dot g(t)|\leq |\dot f(t)|$ for almost all $t\in[a,b]$.
\end{pf}
\section{Proof of Lemma~\ref{lem:TV_formula}}\label{sec:pf:TV}
\begin{pf}
We first show that when $A(t)$	is absolutely continuous over $[a,b]$, its total variation is given by
\begin{equation}\label{formula_AC_TV}
	\int_a^b\Vert dA\Vert=\int_a^b\Vert\dot A(t)\Vert dt.
\end{equation}
Define $\Gamma(t):=\int_a^t\Vert dA\Vert$. Clearly $\Gamma(t)$ is non-decreasing so that when $\dot\Gamma(t)$ exists,
\begin{equation}\label{one_side_for_AC_TV}
	\int_a^b\dot\Gamma(t)dt\leq \Gamma(b)-\Gamma(a)=\Gamma(b)=\int_a^b\Vert dA\Vert.
\end{equation}
On the other hand, it follows from the definition of total variation that $\Vert A(t)-A(s)\Vert\leq\int_s^t\Vert dA\Vert =\Gamma(t)-\Gamma(s)$ for any $t,s\in[a,b],t\geq s$. Thus divide both sides by $t-s$ and take the limit as $t-s\to 0$, we conclude that $\Vert \dot A(t)\Vert\leq \dot\Gamma(t)$ almost everywhere. Combined with \eqref{one_side_for_AC_TV}, we conclude that
\[
\int_a^b\Vert\dot A(t)\Vert dt\leq \int_a^b\Vert dA\Vert.
\]
To show the other opposite inequality, recall that $A(t)$ is assumed to be absolutely continuous over $[a,b]$. From fundamental theorem of Lebesgue integral calculus, we have $A(t_{i+1})-A(t_i)=\int_{t_i}^{t_{i+1}}\dot A(t)dt$ for any $t_i,t_{i+1}\in[a,b],t_i\leq t_{i+1}$. Thus
\[
\Vert A(t_{i+1})-A(t_i)\Vert=\left\Vert\int_{t_i}^{t_{i+1}}\dot A(t)dt\right\Vert\leq \int_{t_i}^{t_{i+1}}\Vert\dot A(t)\Vert dt.
\]
Let $P=\{t_0,t_1,\cdots,t_k\}$ be a partition of $[a,b]$ and appeal to the definition of total variation, we conclude that
\begin{equation}\label{other_side_for_AC_TV}
	\int_a^b\Vert dA\Vert=\sup_{P\in\bP}\sum_{i=1}^k\Vert A(t_{i})-A(t_{i-1})\Vert\leq \int_a^b\Vert\dot A(t)\Vert dt.
\end{equation}
Therefore \eqref{formula_AC_TV} is concluded by combining \eqref{one_side_for_AC_TV} and \eqref{other_side_for_AC_TV}.\\
We now consider the case when $A(t)$ only has one discontinuity at $d\in[a,b]$ and show that
\begin{equation}\label{formula_TV_1_dis}
\int_a^b\Vert dA\Vert=\int_a^d\Vert \dot A(t)\Vert dt+\int_d^b\Vert \dot A(t)\Vert dt +\Vert A(d)-A(d^-)\Vert.
\end{equation}
The proof can be easily adjusted to the case when $A(t)$ has finitely many discontinuities over $[a,b]$, which proves Lemma~\ref{lem:TV_formula}. We first show that
\begin{multline}\label{one_direction}
	\sup_{p\in\mathbb P}\sum_{i=1}^k\Vert A(t_i)-A(t_{i-1}) \Vert\\
	\leq \int_a^d\Vert \dot A(t)\Vert dt+\int_d^b\Vert \dot A(t)\Vert dt +\Vert A(d)-A(d^-)\Vert.
\end{multline} 
To this end, Let $P=\{t_0,t_1,\cdots t_k\}$ be an arbitrary partition of $[a,b]$. Moreover, assume there exists $i^*\in\{1,2\cdots,k\}$ such that $t_{i^*-1}<d\leq t_{i^*}$. Notice that except for $i=i^*$, $A(t)$ is absolutely continuous over $[t_{i-1},t_i]$ so
\begin{equation}\label{ac_pieces}
	\Vert A(t_{i})-A(t_{i-1})\Vert=\Vert\int_{t_{i-1}}^{t_i}\dot A(t)dt\Vert\leq \int_{t_{i-1}}^{t_i}\Vert\dot A(t)\Vert dt. 
\end{equation}
Moreover,
\begin{multline}\label{jump_piece}
	\Vert A(t_{i^*})-A(t_{i^*-1})\Vert\\
	\leq \Vert A(t_{i^*})-A(d)\Vert+\Vert A(d)-A(d^-)\Vert+\Vert A(d^-)-A(t_{i^*-1})\Vert\\
	\leq \int_{d}^{t_{i^*}}\Vert\dot A(t)\Vert dt+\Vert A(d)-A(d^-)\Vert +\int_{t_{i^*-1}}^{d}\Vert\dot A(t)\Vert dt.
\end{multline}
Combining \eqref{ac_pieces} and \eqref{jump_piece}, we conclude that
\begin{multline*}
\sum_{i=1}^k\Vert A(t_i)-A(t_{i-1}) \Vert\\
\leq \int_a^d\Vert \dot A(t)\Vert dt+\int_d^b\Vert \dot A(t)\Vert dt +\Vert A(d)-A(d^-)\Vert.
\end{multline*}
Since the partition $P$ is arbitrary, the above inequality still holds when taking the supremum over $\bP$ and therefore \eqref{one_direction} is shown.\\
We next show the other inequality
\begin{multline}\label{other_direction}
	\sup_{p\in\mathbb P}\sum_{i=1}^k\Vert A(t_i)-A(t_{i-1}) \Vert\\
	\geq \int_a^d\Vert \dot A(t)\Vert dt+\int_d^b\Vert \dot A(t)\Vert dt +\Vert A(d)-A(d^-)\Vert.
\end{multline} 
To this end, let $\epsilon>0$ be arbitrary. because $A(t)$ is absolutely continuous over $[a,d)$ $\int_{s}^d\Vert\dot A(t)\Vert dt$ is continuous with respect to $s$ and there exists $\delta>0$ such that $\Vert A(d^-)-A(d-\delta)\Vert\leq \int_{d-\delta}^{d}\Vert\dot A(t)\Vert dt\leq\frac{\epsilon}{4}$. It can therefore be inferred by triangle inequality that
\begin{multline}\label{other_direction_part_1}
	\Vert A(d)-A(d-\delta)\Vert\geq \Vert A(d)-A(d^-)\Vert-\Vert A(d^-)-A(d-\delta)\Vert \\
	\geq \Vert A(d)-A(d^-)\Vert -\frac{\epsilon}{4}.
\end{multline}
Moreover, since $A(t)$ is absolutely continuous over $[a,d-\delta],[d,b]$, it follows from \eqref{formula_AC_TV} that $\int_a^{d-\delta}\Vert dA\Vert=\int_a^{d-\delta}\Vert\dot A(t)\Vert dt$, $\int_d^b\Vert dA\Vert=\int_d^b\Vert\dot A(t)\Vert dt$, which further means that there exist partitions $P_L=\{t_0,t_1,\cdots, t_{j-1}\}$, $P_R=\{t_j,t_{j+1},\cdots, t_k\}$ of $[a,d-\delta],[d,b]$ respectively, such that
\begin{align}
	\sum_{i=1}^{j-1}\Vert A(t_i)-&A(t_{i-1})\Vert\geq \int_a^{d-\delta}\Vert\dot A(t)\Vert dt-\frac{\epsilon}{4}\nonumber\\
	&=\int_a^{d}\Vert\dot A(t)\Vert dt-\int_{d-\delta}^{d}\Vert\dot A(t)\Vert dt-\frac{\epsilon}{4}\nonumber\\
	&\geq \int_a^{d}\Vert\dot A(t)\Vert dt-\frac{\epsilon}{2} ,\label{other_direction_part_2}\\
	\sum_{i=j+1}^{k}\Vert A(t_i)-&A(t_{i-1})\Vert\geq \int_d^b\Vert\dot A(t)\Vert dt-\frac{\epsilon}{4}.\label{other_direction_part_3}
\end{align}
Combining \eqref{other_direction_part_1},\eqref{other_direction_part_2} and \eqref{other_direction_part_3} together and notice that $t_{j-1}=d-\delta,t_j=d$, we have
\begin{multline*}
\sum_{i=1}^{k}\Vert A(t_i)-A(t_{i-1})\Vert\\
\geq \int_a^d\Vert \dot A(t)\Vert dt+\int_d^b\Vert \dot A(t)\Vert dt +\Vert A(d)-A(d^-)\Vert-\epsilon.
\end{multline*}
Now because $P:=P_L\cup P_R$ is a partition of $[a,b]$ and $\epsilon>0$ is arbitrary, we conclude \eqref{other_direction}. Finally \eqref{formula_TV_1_dis} is concluded by combining \eqref{def:total_variation}, \eqref{one_direction} and \eqref{other_direction}.
\end{pf}
\section{Proof of Lemma~\ref{lem:from_khalil}}\label{sec:pf:P}
\begin{pf}
	Firstly, Assumption~\ref{ass:1} implies $\alpha(A(t))\leq L$. We also have the following two bounds on both the norm of $\tilde A(t)$ and the abscissa of $\tilde A(t)$: $\Vert\tilde A(t)\Vert \leq \Vert A(t)\Vert+\varphi_\kappa(A(t))\leq L+f_{\ramp}(L+\kappa)$, $\alpha(\tilde A(t))=\alpha(A(t))-\varphi_{\kappa}(A(t))\leq -\kappa$. The proof of \eqref{from_khalil_part_1}, \eqref{sandwich_P} follows from the proof of \cite[Lemma 9.9]{Khalil2002}.\\ 
	To show absolute continuity of $P(t)$ and \eqref{bound_derivative_P}, we take any $t,r\in[a,b],t\leq r$. It follows from \eqref{Lyapunov_equation} that
	\begin{align*}
		\tilde A^\top (t)P(t)+P(t)\tilde A(t)&=I,\\
		\tilde A^\top (r)P(r)+P(r)\tilde A(r)&=I.
	\end{align*}
	Take the difference between the two equations, we get
	\begin{multline*}
	\tilde A^\top (t)(P(t)-P(r))+(P(t)-P(r))\tilde A(t)\\
	+	(\tilde A(t)-\tilde A(r))^\top P(r)+P(r)(\tilde A(t)-\tilde A(r))=0.
	\end{multline*}
	Define $\Delta P:=P(t)-P(r),Q:=(\tilde A(t)-\tilde A(r))^\top P(r)+P(r)(\tilde A(t)-\tilde A(r))$. The above equation means that $\Delta P$ is the solution to the Lyapunov equation $\tilde A^\top (t)\Delta P+\Delta P\tilde A(t)+Q=0$.
	Since $\tilde A(t)$ is Hurwitz, the Lyapunov equation has a unique solution given by
	\[
	\Delta P=\int_0^\infty e^{s\tilde A^\top (t)}Qe^{s\tilde A (t)}ds.
	\]
	On one hand we conclude that $\Vert Q\Vert\leq 2\Vert P(r)\Vert\Vert\tilde A(t)-\tilde A(r)\Vert\leq 2c_2\Vert\tilde A(t)-\tilde A(r)\Vert$ by \eqref{sandwich_P}. On the other hand \eqref{from_khalil_part_1} directly gives a bound on $\Vert e^{s\tilde A(t)}\Vert$. Therefore 
	\begin{align*}
		\Vert P(r)- P(t)\Vert&=\Vert\Delta P\Vert\\
		&\leq \int_0^\infty 2\Vert e^{s\tilde A(t)}\Vert\Vert Q\Vert ds \\
		&\leq \int_0^\infty 2cc_2e^{-2\beta s}\Vert\tilde A(t)-\tilde A(r)\Vert ds\\
		&=2c_2^2\Vert\tilde A(t)-\tilde A(r)\Vert.
	\end{align*}
	Appeal to Lemma~\ref{lem:Lipschitz}, we conclude that $P(t)$ is absolutely continuous and \eqref{bound_derivative_P}.
\end{pf}
}
\bibliographystyle{abbrv}
\bibliography{my_ref_library.bib}

\end{document}